\documentclass[12pt]{iopart}

\begin{document}

\title[Space missions to detect the CGB]{Space missions to detect
the cosmic gravitational-wave background}

\author{Neil J. Cornish\footnote[3]{To whom correspondence should be
addressed {\tt (cornish@physics.montana.edu)}} and Shane L. Larson}

\address{Department of Physics, Montana State University, Bozeman,
MT 59717, USA}

\begin{abstract}
It is thought that a stochastic background of gravitational waves was produced
during the formation of the universe. A great deal could be learned by measuring
this Cosmic Gravitational-wave Background (CGB), but detecting the CGB presents a
significant technological challenge.  The signal strength is expected
to be extremely weak, and there will be competition from
unresolved astrophysical foregrounds such as white dwarf binaries.
Our goal is to identify the most promising approach to detect the CGB.
We study the sensitivities that can be reached using both individual, and cross-correlated
pairs of space based interferometers. Our main result is a general, coordinate free
formalism for calculating the detector response that applies to arbitrary detector
configurations. We use this general formalism to identify some promising designs for
a GrAvitational Background Interferometer (GABI) mission. Our conclusion is that
detecting the CGB is not out of reach.

\end{abstract}


%\submitto{\CQG}

\maketitle

\section{Introduction}

Before embarking on a quest to discover the Cosmic Gravitational-wave Background (CGB),
it is worth reflecting on how much has been learned from its electromagnetic
analog, the Cosmic Microwave Background (CMB). The recent
Boomerang\cite{boom} and Maxima\cite{maxima} experiments have furnished
detailed pictures of the universe some 300,000 years after
the Big Bang. These measurements of the CMB anisotropies have been used to infer
that the visible universe is, to a good approximation, spatially flat.
The earlier COBE-DMR\cite{cobe} measurements of the CMB anisotropy
fixed the scale for the density perturbations that formed the large scale structure we see today,
and showed that the density perturbations have an almost scale free spectrum. With all this
excitement over the anisotropy measurements it is easy to forget how much the
CMB taught us before the anisotropies were detected. The initial observation that the
CMB is highly {\em isotropic} is one of the major observational triumphs of the Big
Bang theory. Equally important was the discovery by COBE-FIRAS\cite{cobe2} that the CMB
has an exquisite black body spectrum with a temperature of 2.728 Kelvin.

The lesson that we take from the CMB is that while it would be wonderful to have
a COBE style map of the early universe in gravitational waves, a great deal can
be learned by detecting the isotropic component. The main focus of this work is
on the fixing the amplitude of the CGB at some frequency. Some attention will
also be given to the prospects of measuring the the energy spectrum. Detecting
anisotropies in the CGB is a considerably harder problem that we address
elsewhere\cite{ani}.

\begin{figure}[ht]
\vspace*{3.5in}
\includegraphics{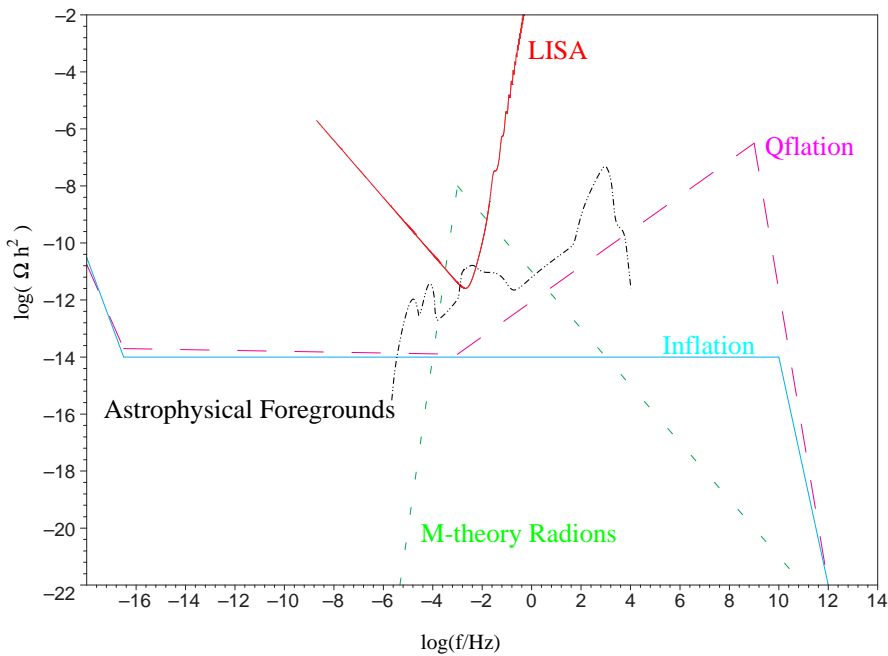}
\caption{Possible sources of a stochastic gravitational wave background plotted
against the sensitivity curve for the LISA mission.}
\end{figure}

There are two major obstacles that stand in the way of detecting the CGB.
The first is the extreme weakness
of the signal, and the second is the competing stochastic background produced by astrophysical
sources. Figure 1 shows various predictions for the CGB energy spectra in several early
universe scenarios including standard inflation, a particular M-theory model\cite{ch} and
quintessence based inflation\cite{qf}. Also shown is the sensitivity curve for the
proposed Laser Interferometer Space Antenna (LISA)\cite{lppa}, and a compilation
of possible extragalactic astrophysical foregrounds taken from the work of
Schneider {\it et al.}\cite{rs}. The spectrum is expressed in terms of $\Omega_{\rm gw}(f)$,
the energy density in gravitational waves (in units of the critical density) per logarithmic
frequency interval, multiplied by $h_0^2$, where $h_0$ is the Hubble constant in units of
100 km s$^{-1}$ Mpc$^{-1}$. The predictions for the CGB power spectra are fairly
optimistic, and some of them come close to saturating existing indirect
bounds on $\Omega_{\rm gw}(f)$ -- see the review by Maggiore for
details\cite{maggiore}.

With the exception of the M-theory motivated model,
we see that the major obstacle to detecting the CGB is not the sensitivity of our
detectors, but competition from astrophysical foregrounds. The astrophysical foregrounds
are produced by the combination of a great number of weak sources that add
together to form a significant stochastic signal. Based on current estimates, it is thought that the
contribution from astrophysical sources falls off fairly rapidly below 1 $\mu$Hz. For
that reason, we will focus our attention on detecting the CGB in the sub-micro Hertz
regime. To put this number into perspective, an interferometer built using LISA technology
would need spacecraft separated by one-tenth of a light year to
reach a peak sensitivity at 1 $\mu$Hz! A more practical approach is to use two smaller
interferometers that lack the intrinsic sensitivity to detect
the CGB on their own, but together are able to reach the necessary sensitivity.
The idea is to cross correlate the outputs from the two interferometers and integrate
over some long observation time $T$. Since the noise in the two interferometers is
uncorrelated while the signal is correlated, the signal to noise ratio will steadily
improve as the observation time is increased.

In principle, it is possible to detect
an arbitrarily weak signal by observing for an arbitrarily long time. In practice, the
prospects are not so rosy as the sensitivity of the detector pair only improves
on the sensitivity of a single detector as $(T\Delta f)^{1/2}$, where $\Delta f$ is
the frequency bandpass. The frequency bandpass is typically taken to equal the
central observing frequency. Suppose we try and look for the CGB at a frequency of
$f\approx \Delta f =10^3$~Hz using the two LIGO detectors. Over an observation time
of $T=1$ year, the
sensitivity is improved by a factor of $\sim 2 \times 10^{5}$ by cross correlating
the Washington and Louisiana detectors. For a pair of LISA detectors with a
bandpass of $\Delta f \approx 10^{-2}$~Hz, cross correlating for a year would
result in a 500 fold improvement in sensitivity. However, for detectors operating in the
$\mu$Hz range, cross correlating two detectors for one year only improves on the sensitivity
of a single detector by a factor of $\sim 6$. So why use two interferometers when one will do?
The reason is simple: with a single interferometer it is not possible to tell the
difference between the CGB and instrument noise since both are, to a good approximation,
stationary Gaussian random processes. Moreover, it is impossible to shield a detector
from gravitational waves in order to establish the noise floor. With two or more
independent interferometers the cross correlated signal can be used to seperate the
stochastic bacground from instrument noise. The LISA system has three arms, and the signal
from these arms can be used to form three different Michelson interferometers. However,
these interferometers share spacecraft, and thus common sources of noise, so nothing
is gained by cross correlating the outputs. However, Tinto {\it et al.}\cite{tae} have recently
suggested another way of combining the signals, known as a Sagnac system, that creates an
interferometer that is fairly insensitive to gravitational waves. Since
the response of this ``bad'' interferometer is noise dominated, its output can be used to
estimate the noise floor for the standard interferometer configuration. Thus, a single
LISA type observatory might be able to discriminate between the CGB and instrument noise.

We will show that cross correlating two LISA interferometers results in a signal to noise
ratio of
\begin{equation}
{\rm SNR} = 1.4 \left(\frac{T}{{\rm year}}\right)^{1/2} \left(\frac{\Omega_{{\rm gw}}
h_0^2}{10^{-14}}\right) 
\end{equation}
for a scale invariant stochastic gravitational wave background. This represents
a five hundred fold improvement over the sensitivity of a single LISA interferometer.
If it were not for the astrophysical foregrounds that are thought to dominate the
CGB signal in the LISA band, a pair of LISA detectors would stand a good chance
of detecting the CGB. In an effort to avoid being swamped by astrophysical
sources, we focus our efforts on detecting the CGB below 1 $\mu$Hz. We show that
a pair of interferometers can reach a signal to noise of
\begin{equation}
    \fl {\rm SNR} = 3.1 \left(\frac{T}{{\rm
    year}}\right)^{1/2} \left(\frac{\Omega_{\rm gw}
    h_{0}^{2}}{10^{-14}}\right)\left(\frac{L}{\sqrt{3} \, {\rm
    AU}}\right)^2 \left(\frac{3\times 10^{-16} {\rm m}{\rm
    s}^{-2}}{\delta a}\right)^2
    \left(\frac{f}{\mu{\rm Hz}}\right)^{3/2} 
\end{equation}
for a scale invariant CGB spectrum. The above result is scaled against a possible
LISA follow-on mission (LISA II) that calls for two
identical interferometers, comprising six equally spaced spacecraft that form two
equilateral triangles
overlayed in a star pattern. The constellations follow circular orbits about
the Sun in a common plane at a radius of 1 AU, so that each interferometer arm is
$L =\sqrt{3}$ AU in length. The acceleration noise, $\delta a$, is scaled against a value
that improves on the LISA specifications by one order of magnitude. The LISA II design
could detect a stochastic background at 99\% confidence with one year of observations if
$\Omega_{\rm gw} h_{0}^{2}=10^{-14}$. Ideally we would like to reach a sensitivity of at
least $\Omega_{\rm gw}h_{0}^{2} \sim 10^{-20}$ in order to detect inflationary spectra with
a mild negative tilt. It is difficult to achieve this with spacecraft orbiting at 1 AU as
thermal fluctuations due to solar heating of the spacecraft make it hard to reduce the
acceleration noise below $\delta a \simeq 10^{-16}\, {\rm m}{\rm s}^{-2}$.
The best strategy is to increase the size of the orbit, $R$, as this
increases the size of the interferometer, $L = \sqrt{3} R$, and decreases the
thermal noise by $R^{-2}$.

\section{Outline}

We begin by deriving the response of a pair of cross correlated space-based laser
interferometers to a stochastic background of gravitational waves. Similar calculations
have been done for ground based detectors\cite{mich,christ,flan,bruce,ar}, but
these omit the high frequency transfer
functions that play an important role in space based systems. Our calculations closely follow those
of Allen and Romano\cite{ar}, and reduce to theirs in the low
frequency limit\footnote{Here high and low frequencies
are defined relative to the {\em transfer frequency} of the detector $f_* = c/(2\pi L)$,
where $L$ is the length of one interferometer arm.}.
Many of the calculational details that we omit for brevity can be found in their paper.
Having established the general formalism for cross correlating space based interferometers
we use our results to identify some promising configurations for a GrAvitational Background
Interferometer (GABI) mission to detect the CGB.

\section{Detector Response}

We attack the problem of cross correlating two space based
interferometers in stages. We begin by reviewing how the Doppler
tracking of a pair of spacecraft can be used to detect gravitational waves.
Using this result we derive the response of a two arm Michelson interferometer
and express the result in a convenient coordinate-free form. The response of
a single interferometer is then used to find the sensitivity that can be
achieved by cross correlating two space based interferometers.

\subsection{Single arm Doppler tracking}

Following the treatment of Hellings\cite{ron}, we consider two free spacecraft,
one at the origin of a coordinate system and the other a distance
$L$ away at an angle $\theta$ from the $z$ axis.  Both spacecraft are
at rest in this coordinate system.  The spacecraft at the origin sends
out a series of photons, while a weak plane gravitational wave is
passing through space in the $+z$ direction.  To leading order, the
spacetime metric in the transverse-traceless gauge is:
\begin{equation}
    \fl \quad ds^2 = - c^2 dt^2 + dz^2 + (1+h\cos2\psi)dx^2 
    + (1-h\cos2\psi)dy^2 - h\sin2\psi dxdy \, ,
\end{equation}
where $h(t-z/c)$ is the wave amplitude and $\psi$ is the angle between
the principal polarization vector and the $x$ axis.  If we choose our
coordinates such that the second spacecraft is in the $x-z$ plane, the
path of the photons can be parameterized by
\begin{equation}
   x=\rho \sin\theta \quad \quad z=\rho \cos\theta \, .
\end{equation}
The photon path in the perturbed spacetime is given by $ds^2=0$, or
\begin{equation}
   c\, dt \simeq \pm (1+\frac{1}{2}h\cos 2\psi\sin^2\theta)\, d\rho \, .
\end{equation}
The round trip journey from spacecraft 1 to spacecraft 2 and back
again is given by
\begin{eqnarray}
   \fl \quad \ell(t_2-t_0)=\int_{t_0}^{t_2} c\, dt &=& 2L +
   \frac{1}{2}\cos2\psi\sin^2\theta \left( \int_0^{L}
   h[t_0+\frac{\rho}{c}(1-\cos\theta)]d\rho \right.  \nonumber \\
   &&\left.  \hspace*{1in}+\int_0^{L}
   h[t_1+\frac{L}{c}-\frac{\rho}{c}(1+\cos\theta)]d\rho\right)
   \, .
\end{eqnarray}
Here $t_0$ is the time of emission, $t_1$ is the time of reception at
spacecraft 2 and $t_2$ is the time of reception back at the first
spacecraft.  Using $h(q)=h_0\exp(i\omega q)$, the varying portion of
the round-trip distance is
\begin{eqnarray}\label{ron}
    \fl \delta\ell(t_2)=L\, h(t_2)\cos2\psi\sin^2\theta
    \frac{1}{2}&&\left( {\rm sinc}\left(\frac{f}{2f_*
    }(1-\cos\theta)\right) \exp\left(-i\frac{f}{2f_*}(3
    +\cos\theta)\right)\right.  \nonumber \\
    && \left.  +{\rm sinc}\left(\frac{f}{2f_* }(1+\cos\theta)\right)
    \exp\left(-i\frac{f}{2f_*}(1+\cos\theta)\right)\right) \, ,
\end{eqnarray}
where $f=\omega/(2\pi)$ is the frequency of the gravitational wave and
$f_*=c/(2\pi L)$ is the {\em transfer frequency}. To zeroth order in $h$,
the optical path length is $\ell = 2L$. The general, coordinate
independent version of this expression is given by
\begin{equation}\label{one}
    \frac{\delta\ell}{\ell} = \frac{1}{2}({\bf u}\otimes{\bf u}):{\bf h}(\widehat\Omega)
    \, {\cal T}({\bf u}\cdot\widehat{\Omega},f) \, ,
\end{equation}
where $\bf{u}$ is a unit vector pointing from the first to the second spacecraft
and $\widehat\Omega$ is a unit vector in
the direction the gravitational wave is propagating.  The colon
denotes the double contraction ${\bf a}:{\bf b}=a_{ij}b^{ij}$.  The
transfer function ${\cal T}$ is given by
\begin{eqnarray}
    {\cal T}({\bf u}\cdot\widehat{\Omega},f)&=&\frac{1}{2}\left[
    {\rm sinc}\left( \frac{f}{2f_* }(1-{\bf
    u}\cdot\widehat{\Omega})\right)\exp\left(-i\frac{f}{2f_*}(3 +{\bf
    u}\cdot\widehat{\Omega})\right) \right. \nonumber \\
    &+& \left. {\rm sinc}\left(\frac{f}{2f_* }(1+{\bf
    u}\cdot\widehat{\Omega})\right) \exp\left(-i\frac{f}{2f_*}(1+{\bf
    u}\cdot\widehat{\Omega})\right)\right] \, .
\end{eqnarray}
This expression for ${\cal T}$ agrees with the one derived by Schilling\cite{sch}.
The gravitational wave is described by the tensor
\begin{equation}
    {\bf h}(t,{\bf x}) = h^+(\omega t -\omega\widehat{\Omega}\cdot{\bf
    x}){\bf \epsilon}^+(\widehat{\Omega},\psi)+
    h^\times(\omega t-\omega \widehat{\Omega}\cdot{\bf x}) {\bf
    \epsilon}^\times(\widehat{\Omega},\psi) \, ,
\end{equation}
with polarization tensors
\begin{eqnarray}
    &&{\bf \epsilon}^+(\widehat{\Omega},\psi)={\bf
    e}^+(\widehat{\Omega})\cos 2\psi- {\bf
    e}^\times(\widehat{\Omega})\sin 2\psi \nonumber \\
    &&{\bf \epsilon}^\times(\widehat{\Omega},\psi)={\bf
    e}^+(\widehat{\Omega})\sin 2\psi+ {\bf
    e}^\times(\widehat{\Omega})\cos 2\psi \, .
\end{eqnarray}
The basis tensors can be written as
\begin{eqnarray}
    &&{\bf
    e}^+(\widehat{\Omega})=\hat{m}\otimes\hat{m}-\hat{n}\otimes\hat{n}
    \nonumber \\
    &&{\bf
    e}^\times(\widehat{\Omega})=\hat{m}\otimes\hat{n}+\hat{n}\otimes\hat{m}
\end{eqnarray}
where $\hat{m}$, $\hat{n}$ and $\widehat{\Omega}$ are an orthonormal set of
unit vectors. The expression in (\ref{ron}) can be recovered from (\ref{one}) by setting
\begin{eqnarray}
&& h^+ =h, \quad h^\times =0, \quad \widehat\Omega = \hat{z}, \quad
{\bf u}=\hat{x}\sin\theta+\hat{z}\cos\theta, \nonumber \\
&& {\bf e}^+=\hat{x}\otimes\hat{x}-\hat{y}\otimes\hat{y}, \;\; {\rm and} \; \;
{\bf e}^\times = \hat{x}\otimes\hat{y}+\hat{y}\otimes\hat{x}\, .
\end{eqnarray}

The transfer function ${\cal T}$ approaches unity for $f \ll f_*$, and
falls of as $1/f$ for $f \gg f_*$ by virtue of the sinc function.  For
a ground based detector such as LIGO the transfer function can be ignored
since instrument noise keeps the operational range below LIGO's transfer
frequency of $f_* \approx 10^4$ Hz (not to mention that LIGO is a Fabry-Perot
interferometer so our results do not really apply\cite{christ}). For a space based
detector such as LISA, the instrument noise does not rise appreciably
at high frequencies, and it is the transfer function that limits the
high frequency response.  The sinc function is
responsible for the wiggly rise in the LISA sensitivity
curve at high frequency that can be seen in Figure 4.

\subsection{Interferometer response}

We can form an interferometer by introducing a third spacecraft at a
distance $L$ from the corner spacecraft and differencing the outputs
of the two arms\footnote{More complicated differencing schemes have to
be applied if the arms of the interferometer have unequal lengths in
order to cancel laser phase noise\cite{uneq}.}.  The measured strain is given by
\begin{eqnarray}\label{inter}
     s(\widehat\Omega,f,{\bf x},t) &=& \frac{ \delta \ell_{{\bf u}}(t) - \delta \ell_{{\bf
    v}}(t)}{\ell} \nonumber \\ 
     &=&{\bf D}(\widehat\Omega,f):{\bf h}(\widehat\Omega,f,{\bf x},t) \, ,
\end{eqnarray}
where
\begin{equation}\label{dresp}
{\bf D}(\widehat\Omega,f) = \frac{1}{2}\left( ({\bf u}\otimes{\bf u})
        \, {\cal T}({\bf u}\cdot\widehat{\Omega},f) -
    ({\bf v}\otimes{\bf v})\, {\cal T}({\bf
    v}\cdot\widehat{\Omega},f) \right) 
\end{equation}
is the detector response tensor and
${\bf u}$ and ${\bf v}$ are unit vectors in the direction of
each interferometer arm, directed out from the vertex of the interferometer.
The above expression gives the response of the
interferometer to a plane wave of frequency $f$ propagating in the $\widehat\Omega$
direction. A general gravitational wave background can be expanded in the terms of
plane waves:
\begin{eqnarray}\label{sto}
    h_{ij}(t,{\bf x}) &=&
    \int_{-\infty}^{\infty} df \, \int d\widehat{\Omega} \;
    \tilde{h}_{ij}(\widehat\Omega,f,{\bf x},t) \nonumber \\
    &=& \sum_{A}
    \int_{-\infty}^{\infty} df \int d\widehat{\Omega} \;
    \tilde{h}_A(f,\widehat{\Omega})e^{-2\pi i ft}e^{2\pi i f
    \widehat{\Omega}\cdot{\bf x}/c}
    e_{ij}^{A}(\widehat{\Omega}) \, .
\end{eqnarray}
Here $ \int d\widehat{\Omega} =\int_0^{2\pi} d\phi \int_{0}^{\pi} \sin\theta d\theta$
denotes an all sky integral.
The Fourier amplitudes obey $\tilde{h}_A(-f)=\tilde{h}^*_A(f)$ since the waves have
real amplitudes. In the final expression we have chosen $\bf{e}^{+}$ and
$\bf{e}^{\times}$ as basis tensors for the decomposition of the two independent
polarizations. The response of the detector (which we are free to locate at
$\bf{x}={\bf 0}$) to a superposition of plane waves is
then
\begin{eqnarray}
s(t) &=& \int_{-\infty}^{\infty} df \, \int d\widehat{\Omega} \;
s(\widehat\Omega,f,{\bf 0},t) \nonumber \\
&=& \sum_{A} \int_{-\infty}^{\infty} df  \, \int d\widehat{\Omega} \;
\tilde{h}_A(f,\widehat{\Omega})e^{-2\pi i ft}\,
{\bf D}(\widehat\Omega,f):{\bf e}^{A}(\widehat\Omega) .
\end{eqnarray} From this we infer that the
Fourier transform of $s(t)$ is given by
\begin{equation}
\tilde{s}(f) = \sum_{A} \int d\widehat{\Omega} \;
    \tilde{h}_A(f,\widehat{\Omega})
    {\bf D}(\widehat\Omega,f):{\bf e}^{A}(\widehat\Omega) \, .
\end{equation}
In the low frequency limit the transfer function ${\cal
T}$ approaches unity and ${\bf D}(\widehat\Omega,f)$ only depends on the geometry of the detector:
\begin{equation}
    {\bf D} =\frac{1}{2} \left( {\bf u}\otimes{\bf u} - {\bf
    v}\otimes{\bf v} \right) \, .
\end{equation}

To a first approximation, a stochastic gravitational wave background can be taken
to be isotropic, stationary and unpolarized. It is fully specified by the
ensemble averages:
\begin{equation}\label{ens1}
\langle \tilde{h}_A(f,\widehat{\Omega}) \rangle =0 \; , \quad  \quad
\langle \tilde{h}_A(f,\widehat{\Omega})
    \tilde{h}_{A'}(f',\widehat{\Omega}')\rangle = 0 \; ,
\end{equation}
and
\begin{equation}\label{ens}
    \langle
    \tilde{h}^*_A(f,\widehat{\Omega})
    \tilde{h}_{A'}(f',\widehat{\Omega}')\rangle
    =\frac{1}{2}\delta(f-f')
    \frac{\delta^2(\widehat{\Omega},\widehat{\Omega}')}
    {4\pi}\delta_{AA'} \, S_h(f) \, .
\end{equation}
Here $S_h(f)$ is the spectral density of the stochastic background and
the normalization is chosen such that
\begin{eqnarray}
\langle \tilde{h}^*_A(f)\tilde{h}_{A'}(f')\rangle &\equiv &
\int d\widehat{\Omega}\, d\widehat{\Omega}'
\langle \tilde{h}^*_A(f,\widehat{\Omega})\, 
    \tilde{h}_{A'}(f',\widehat{\Omega}')\rangle \nonumber \\
&=& \frac{1}{2}\delta(f-f')\delta_{AA'} \, S_h(f) \, .
\end{eqnarray}
The spectral
density has dimension Hz$^{-1}$ and satisfies $S_h(f)=S_h(-f)$. 
It is related to $\Omega_{\rm gw}(f)$ by
\begin{equation}\label{ogw}
    \Omega_{\rm gw}(f) = \frac{4 \pi^2}{3H_0^2} f^3 S_h(f) \, .
\end{equation}
The time averaged response\footnote{The limit $\tau \rightarrow \infty$ is
approximated in practice by observing for a period much longer than
the period of the gravitational wave.} of the interferometer,
\begin{equation}
   \langle s(t)\rangle _\tau = \lim_{\tau \rightarrow \infty} \frac{1}{\tau }
   \int_{-\tau/2}^{\tau /2} s(t) dt \, ,
\end{equation}
when applied to a stochastic background, is equivalent to the ensemble average
\begin{eqnarray}
\langle s(t)\rangle _\tau &=& \langle s(t)\rangle \nonumber \\
&=& \sum_{A} \int_{-\infty}^{\infty} df  \, \int d\widehat{\Omega} \;
\langle \tilde{h}_A(f,\widehat{\Omega})\rangle e^{-2\pi i ft}\,
{\bf D}(\widehat\Omega,f):{\bf e}^{A}(\widehat\Omega) \nonumber \\
&=& 0 \, .
\end{eqnarray}
Since the expectation value of
$s(t)$ vanishes we need to consider higher moments such as $s^2(t)$.
Using equations
(\ref{inter}) and (\ref{sto}) we find
\begin{equation}
   \langle  s^2(t)\rangle = \int_0^\infty df S_h(f) {\cal R}(f) \, ,
\end{equation}
where the transfer function ${\cal R}(f)$ is given by
\begin{equation}\label{rtran}
     {\cal R}(f) = \int \frac{d \widehat{\Omega}}{4\pi} \sum_{A}
    F^A(\widehat{\Omega},f)F^A(\widehat{\Omega},f)^* \, ,
\end{equation}
and $F^A(\widehat{\Omega},f)={\bf D}(\widehat{\Omega},f):{\bf e}^{A}(\widehat{\Omega})$
is the detector response function. 
In the low frequency limit, $f\ll f_*$, it is easy to show that ${\cal R}(f)
= 2/5\, \sin^2\beta$, where $\cos\beta = {\bf u}\cdot{\bf v}$ is the
angle between the interferometer arms.  

The response of the interferometer can be expressed in terms of the {\em strain spectral
density}, $\tilde{h}_s(f)$, which has units of Hz$^{-1/2}$ and is defined by
\begin{equation}\label{spow}
 \langle  s^2(t)\rangle  = \int_0^\infty df \tilde{h}_s^2(f) = \int_0^\infty df S_h(f)
{\cal R}(f) 
\end{equation}
We see that the strain spectral density in the interferometer is related to the spectral
density of the source by
\begin{equation}
 \tilde{h}_s(f) = \sqrt{S_h(f) {\cal R}(f)} \, .
\end{equation}
The total output of the interferometer is given by the sum of the signal and the noise:
\begin{equation}
S(t) = s(t) + n(t) \, .
\end{equation}
Assuming the noise is Gaussian, it can be fully characterized by
the expectation values
\begin{equation}
\langle \tilde{n}(f) \rangle =0 \, , \quad {\rm and} \quad
\langle \tilde{n}^*(f) \tilde{n}(f')\rangle = \frac{1}{2}\delta(f-f')S_n(f) \, ,
\end{equation}
where $S_n(f)$ is the noise spectral density.
The total noise power in the interferometer is thus
\begin{equation}\label{npow}
\langle n^2(t) \rangle = \int_0^\infty df S_n(f) = \int_0^\infty df \, \tilde{h}_n^2(f)
\end{equation}
where $\tilde{h}_n(f)$ is the the strain spectral density due to the noise. Comparing
equations (\ref{spow}) and (\ref{npow}), we define
the signal to noise ratio at frequency $f$ by
\begin{equation}\label{snrf}
{\rm SNR}(f) = \frac{\tilde{h}_s^2(f)}{\tilde{h}_n^2(f)} = \frac{S_h(f) {\cal R}(f)}{S_n(f)} \, .
\end{equation}
Sensitivity curves for space-based interferometers typically display the effective
strain noise
\begin{equation}\label{hef}
\tilde{h}_{\rm eff}(f) = \sqrt{\frac{S_n(f)}{{\cal R}(f)}} \, .
\end{equation}
At high frequencies it is the decay of the transfer function ${\cal R}(f)$ leads to
a rise in the effective noise floor. The actual noise power, $S_n(f)$, does not
rise significantly at high frequencies for space based systems.

\subsection{The LISA interferometer}

\begin{figure}[ht]
\vspace*{3.3in}
\includegraphics{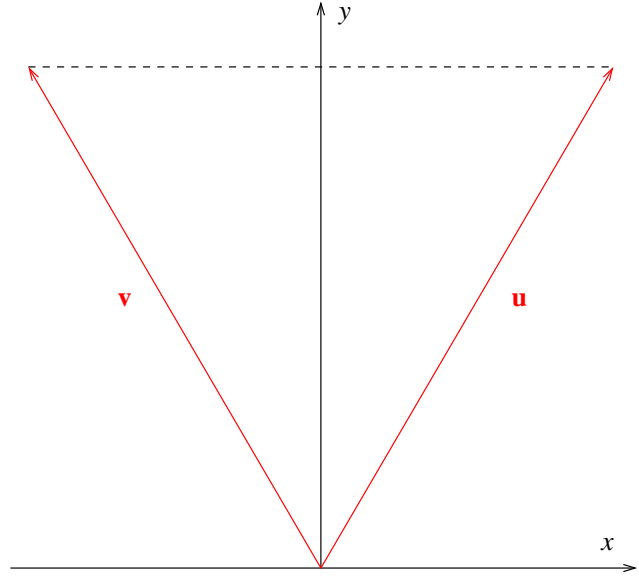}
\caption{The coordinate system used to evaluate LISA's transfer function.}
\end{figure}

As a concrete example of the general formalism described above we derive the
sensitivity curve for the LISA interferometer using the design specifications
quoted in the LISA Pre-Phase A Report\cite{lppa}. Using the coordinate system
shown in Figure 2, the unit vectors along each arm are given by
\begin{eqnarray} 
    &&{\bf u}
    =\frac{1}{2}\hat{x}+\frac{\sqrt{3}}{2}\hat{y} \nonumber \\
    &&{\bf v} = -\frac{1}{2}\hat{x}+\frac{\sqrt{3}}{2}\hat{y} \, ,
\end{eqnarray}
and the gravitational wave is described by
\begin{eqnarray}
    && \widetilde{\Omega} = \cos\phi \, \sin\theta \, \hat{x} +
    \sin\phi \, \sin\theta \, \hat{y} + \cos\theta \, \hat{z} \\
    && \hat{m} = \sin\phi \, \hat{x} -\cos\phi \, \hat{y} \\
    && \hat{n} = \cos\phi \, \cos\theta\, \hat{x} + \sin\phi \,
    \cos\theta\, \hat{y} -\sin\theta \, \hat{z} \, .
\end{eqnarray}
The angle between the interferometer arms is ${\rm arccos}({\bf u}\cdot{\bf v})=\beta=\pi/3$.
The various ingredients we need to calculate ${\cal R}(f)$ are
\begin{eqnarray}\label{ing1}
   &&{\bf u}\cdot \widehat{\Omega} = \sin(\phi+\pi/6)\sin\theta \\
   &&{\bf v}\cdot \widehat{\Omega} = \sin(\phi-\pi/6)\sin\theta 
\end{eqnarray}
and
\begin{equation}
    ({\bf u}\otimes{\bf u}):{\bf e}^+ = \frac{1}{4}\sin^2\theta
    +\frac{1}{2} \cos^2\theta\cos^2\phi -\frac{\sqrt{3}}{4}\sin
    2\phi\, (1+\cos^2\theta) \, .
\end{equation}

\begin{equation}
    ({\bf v}\otimes{\bf v}):{\bf e}^+ = \frac{1}{4}\sin^2\theta
    +\frac{1}{2} \cos^2\theta\cos^2\phi +\frac{\sqrt{3}}{4}\sin
    2\phi\, (1+\cos^2\theta) \, .
\end{equation}

\begin{equation}
   ({\bf u}\otimes{\bf u}):{\bf e}^\times = -\cos\theta
   \sin(2\phi+\pi/3) \, .
\end{equation}

\begin{equation}\label{ing2}
   ({\bf v}\otimes{\bf v}):{\bf e}^\times = -\cos\theta
   \sin(2\phi-\pi/3) \, .
\end{equation}

Before proceeding to find ${\cal R}(f)$ it is instructive to evaluate the
detector response functions, $F^A(\widehat{\Omega},f)$, at zero frequency:
\begin{eqnarray}
F^{+}(\widehat{\Omega}) &=& \frac{1}{2}({\bf u}\otimes{\bf u}-{\bf v}\otimes{\bf v}):{\bf e}^+ \nonumber \\
&=& -\frac{\sqrt{3}}{4}\sin 2\phi\, (1+\cos^2\theta) \, ,
\end{eqnarray}
and
\begin{eqnarray}
F^{\times}(\widehat{\Omega}) &=& \frac{1}{2}({\bf u}\otimes{\bf u}
-{\bf v}\otimes{\bf v}):{\bf e}^\times \nonumber \\
&=& -\frac{\sqrt{3}}{2}\cos 2\phi\, \cos\theta\, ,
\end{eqnarray}
The magnitudes of these response functions are shown in Figure 3. They can be thought of
as the polarization dependent antenna patterns for the LISA detector appropriate to a
stochastic background of gravitational waves\footnote{These shapes are not coordinate independent.
They depend on how we choose ${\bf u}$ and ${\bf v}$ in relation to ${\bf e}^+$ and
${\bf e}^\times$}.

\begin{figure}[ht]
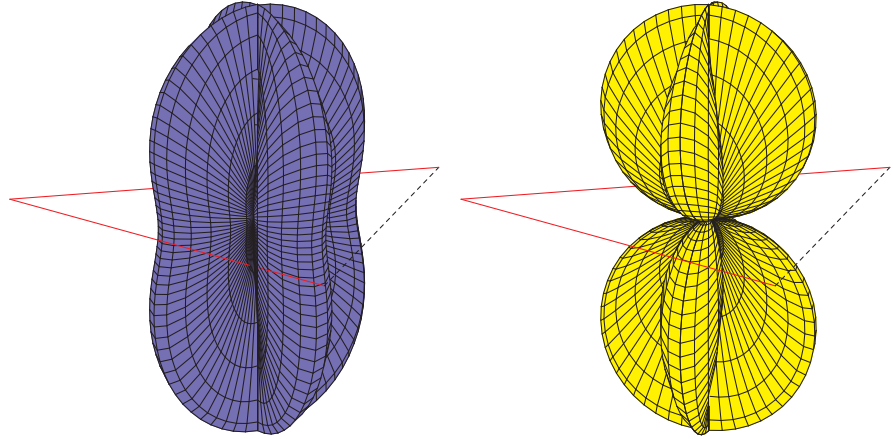

\vspace*{2.9in}
\includegraphics{Fp.ps}
\includegraphics{Fc.ps}
\caption{The magnitudes of the detector response functions $F^+(\widehat{\Omega},f)$ and
$F^\times(\widehat{\Omega},f)$ in the low frequency limit.}
\end{figure} 

Inserting equations (\ref{ing1}) through (\ref{ing2}) into equation (\ref{rtran}) gives
the transfer function ${\cal R}(f)$ in terms of an integral over the angles $\theta$ and
$\phi$. Our explicit expression for LISA's transfer function agrees with the one
found by Larson, Hiscock and Hellings\cite{lhh} using an alternative approach.
We were unable to perform the angular integral to arrive at a general form for ${\cal R}(f)$,
but for low frequencies the integrand can be expanded in a
series to give
\begin{eqnarray}
     && {\cal R}(f) = \frac{2}{5}\sin^2\beta \left( 1 -
    \frac{37\cos^2(\beta/2)-10\cos^4(\beta/2)-1}{84
    \cos^2(\beta/2)}\left(\frac{f}{f_*}\right)^2 
    \right. \nonumber \\
     && + \left. 
    \frac{20\cos^6(\beta/2)+163\cos^2(\beta/2)-80\cos^4(\beta/2)-6}
    {2268\cos^2(\beta/2)} \left(\frac{f}{f_*}\right)^4 - \dots \right).
\end{eqnarray}
At frequencies above $f \sim \frac{3}{2} f_*$, the sinc function takes over and the
transfer function falls of as $1/f^2$.
Setting $\beta=\pi/3$, we find that a good approximation for the transfer function
is given by
\begin{equation}\label{rapp}
 \fl   {\cal R}(f)=\left \{\begin{array}{ll} {\displaystyle \frac{3}{10}\left(
    1- \frac{169}{504}\left(\frac{f}{f_*}\right)^2
    + \frac{425}{9072}
    \left(\frac{f}{f_*}\right)^4 - \frac{165073}{47900160} \left(\frac{f}{f_*}\right)^6
    \right)} ,\quad & f
    < \frac{3}{2}f_* \\ \bs
    {\displaystyle \frac{16783143}{126156800}\left(\frac{3f_*}{2f}\right)^2}, 
     \quad & f\geq \frac{3}{2} f_*
    \end{array} \right.  .
\end{equation}
The coefficient in front of the high frequency term is chosen so that the transfer
function is continuous at $f= \frac{3}{2} f_*$.

\begin{figure}[ht]
\vspace*{3.5in}
\includegraphics{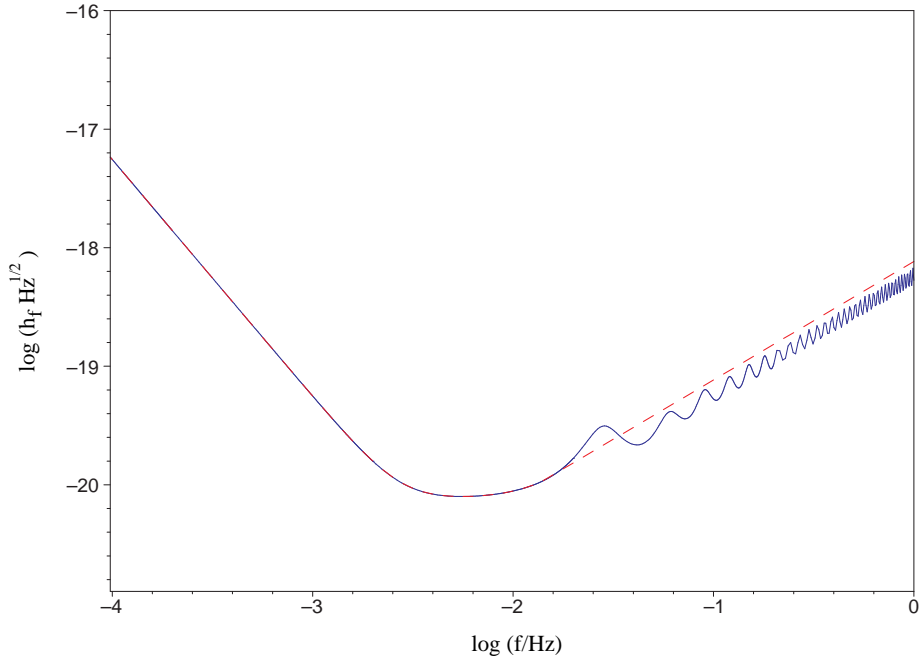}
\caption{The effective noise floor for the LISA mission. The solid line was obtained
numerically while the dashed line is our analytic approximation.}
\end{figure} 

Now that we have calculated the transfer function, our next task is to estimate the
the detector noise power, $S_n(f)$. There are many noise contributions discussed
in the LISA Pre-Phase A Report\cite{lppa}. The dominant ones are thought to be
acceleration noise from the inertial sensors and position noise due to laser shot noise.
A position noise of $\widetilde{\delta x}= 2 \times 10^{-11}$ m Hz$^{-1/2}$ is quoted for each
LISA spacecraft. There are two such contributions per arm, giving a total
of 4 contributions for the interferometer. Since the contributions are uncorrelated they
add in quadrature to give a total position noise of $2\widetilde{\delta x}$. Dividing
this by the optical path length of $2L$ and squaring gives the position noise
power:
\begin{equation}
S_n^{\rm pos}(f) = \left(\frac{\widetilde{\delta x}}{L}\right)^2 \; \, .
\end{equation}
An acceleration noise of $\widetilde{\delta a} = 3 \times 10^{-15}$ ms$^{-2}$  Hz$^{-1/2}$
is quoted for each inertial sensor. This noise acts coherently on the incoming and
outgoing signal, for a combined acceleration noise of $2\widetilde{\delta a}$ per
spacecraft. There are
four such contributions in the interferometer that add in quadrature for a total
acceleration noise of $4\widetilde{\delta a}$. Dividing this by the square of the
angular frequency of the gravitational wave yields the effective position noise
due to spurious accelerations. Dividing the effective position noise by the optical
path length and squaring gives the acceleration noise power:
\begin{equation}\label{accel}
S_n^{\rm accl}(f) = \left(\frac{2\widetilde{\delta a}}{(2\pi f)^2 L}\right)^2 \, .
\end{equation}
Adding together the acceleration and position noise power gives the total noise power.
Using the nominal values for the LISA mission\cite{lppa} we obtain
\begin{equation}\label{lisan}
S_n(f) =\left(9.24\times 10^{-40} \left(\frac{{\rm mHz}}{f}\right)^4 
+ 1.6\times 10^{-41} \right) \; {\rm Hz}^{-1} \, .
\end{equation}
Inserting the above expression for $S_n(f)$, along with  
the approximate expression for the transfer function (\ref{rapp}),
into equation (\ref{hef}) yields a useful analytic approximation
for the effective strain noise, $\tilde{h}_{\rm eff}(f)$, in the LISA
interferometer. The analytic approximation is compared
to the full numerical result in Figure 4.

LISA reaches a peak sensitivity in the frequency range $3 \times 10^{-3} \rightarrow
 10^{-2}$ Hz. Using equations (\ref{snrf}) and (\ref{hef}) we have
\begin{equation}
S_h(f) = \tilde{h}_{\rm eff}^2(f) \, {\rm SNR}(f) \, .
\end{equation}
For a SNR of 2, this translates into a sensitivity of
\begin{equation}
\tilde{h}_{\rm eff}(f) = 6.2 \times 10^{-21}
\left(\frac{\Omega_{\rm gw}(f) h_0^2}{10^{-13}}\right)^{1/2}
\left(\frac{f}{{\rm mHz}}\right)^{3/2} \, {\rm Hz}^{-1/2}\, .
\end{equation}
Thus, a stochastic background with $\Omega_{\rm gw}(f) h_0^2 > 7\times 10^{-12}$ should
dominate LISA's instrument noise for frequencies near 3 mHz. The difficulty would
come in deciding if the interferometer response was due to instrument noise or a
stochastic background, as both are Gaussian random processes. One way to be sure
is to fly two LISA interferometers and cross correlate their outputs.

\section{Cross correlating two interferometers}

Having demonstrated that our formalism recovers the standard results for a single
interferometer, we now investigate how the sensitivity can be improved by
cross correlating two interferometers.
The most general form for the cross correlation of two
detectors is given by
\begin{eqnarray}
    C &=& \int_{-T/2}^{T/2} dt \int_{-T/2}^{T/2} dt' \, S_1(t)S_2(t')
    Q(t-t') \nonumber \\
    &\simeq & \int_{-\infty}^{\infty} df \int_{-\infty}^{\infty} df'
    \delta_T(f-f') \tilde{S}_1^*(f) \tilde{S}_2(f') \widetilde{Q}(f')\, .
\end{eqnarray}
Here $Q(t-t')$ is a filter function and the
$S_i$'s are the strain amplitudes that we read out from the
$i^{\rm th}$ detector:
\begin{equation}
    S_i(t) = h_i^{\rm astro}(t)+s_i(t)+n_i(t) \, .
\end{equation}
The contributions are from resolvable astrophysical sources $h_i^{\rm
astro}$, the stochastic background $s_i$, and intrinsic detector noise
$n_i$. The function $\delta_T$ that appears in the Fourier space
version of the correlation function is the ``finite time delta function''
\begin{equation}
    \delta_T = \int_{-T/2}^{T/2} dt \, e^{-2\pi i ft} = \frac{\sin(\pi f T)}{\pi f} \, .
\end{equation}
It obeys
\begin{equation}
\delta_T(0) = T \, \quad {\rm and} \quad \delta(x) = \lim_{T\rightarrow \infty}\delta_T(x) \, .
\end{equation}

After removing  resolvable astrophysical sources, the ensemble
average of $C$ is given by
\begin{equation}
    \langle C\rangle  = \langle s_1,s_2\rangle 
    +\langle s_1,n_2\rangle +\langle n_1,s_2\rangle
    +\langle n_1,n_2\rangle  =  \langle s_1,s_2\rangle \, .
\end{equation}
Notice that terms involving the noise vanish as the noise is uncorrelated
with the signal, and the noise in different detectors is also uncorrelated. Thus,
\begin{equation}\label{sig}
    \langle C\rangle  = \langle s_1,s_2\rangle  = \int_{-\infty}^{\infty} df
    \int_{-\infty}^{\infty} df' \delta_T(f-f') \langle \tilde{s}_1^*(f)
    \tilde{s}_2(f')\rangle \, \widetilde{Q}(f') \, .
\end{equation}
Our next task is to evaluate the quantity $\langle\tilde{s}_1^*(f) 
\tilde{s}_2(f')\rangle$ that appears in equation (\ref{sig}). Taking the Fourier
transform of the expression in (\ref{inter}) and performing the
ensemble average we find
\begin{equation}\label{sc}
    \langle\tilde{s}_1^*(f) \tilde{s}_2(f')\rangle =
    \frac{1}{2}\delta(f-f')S_h(f)\,  \gamma(f)\, \frac{2}{5}\sin^2\beta\, ,
\end{equation}
where $\gamma(f)$ is the overlap reduction function
\begin{equation}\label{overlap}
    \fl \gamma(f) = \frac{5}{2\sin^2\beta}\int \frac{d
    \widehat{\Omega}}{4\pi}
    \left({F_1^{+}}^*(\widehat{\Omega},f)F_2^{+}(\widehat{\Omega},f)+
    {F_1^{\times}}^*(\widehat{\Omega},f)F_2^{\times}(\widehat{\Omega},f)\right)
    e^{-2\pi i f \widehat{\Omega}\cdot({\bf x}_1-{\bf x}_2)/c} \, .
\end{equation}
The normalization is chosen so that $\gamma(0)=1$. The function $\gamma(f)$ is
Hermitian in the sense that $\gamma_{12}(f)=\gamma^*_{21}(f)$. The inverse
Fourier transform of $\gamma(f)$ is a real function since $\gamma^*(f)=\gamma(-f)$.
The term ``overlap reduction function'' refers to the fact that $\gamma(f)$ takes into
account the misalignment and separation of the interferometers.
In addition, $\gamma(f)$ contains the transfer functions
for each detector, as can be
seen by setting $F_1 = F_2$ in (\ref{overlap}) and comparing with
(\ref{rtran}):
\begin{equation}\label{coco}
\gamma_{1=2}(f) = \frac{5}{2 \sin^2\beta}\;  {\cal R}(f)\, .
\end{equation}
In other words, $\gamma(f)$ is an overlap reduction function and
detector transfer function all rolled into one. Inserting equation (\ref{sc}) into
equation (\ref{sig}) yields the expectation value of $C$:
\begin{eqnarray}\label{sigx}
    \langle C\rangle  &=& \frac{T}{5} \sin^2\beta\, \int^{\infty}_{-\infty} df
    \, S_h(f) \gamma(f) \widetilde{Q}(f) \nonumber \\
                &=& \frac{2T}{5} \sin^2\beta\, \int^{\infty}_0 df
    \, S_h(f) \Re [\gamma(f) \widetilde{Q}(f)] \, .
\end{eqnarray}
In the final line we have switched from the mathematically convenient range of integration
$f \in (-\infty,\infty)$ to the physically relevant range $f\in [0,\infty)$.

The noise in a measurement of $C$ is given by $N=C-\langle C\rangle$,
and the signal to noise ratio (squared) is given by
\begin{equation}
{\rm SNR}^2 \equiv \frac{\langle C\rangle^2 }
    {\langle N^2 \rangle} = \frac{\langle C\rangle^2 }
    {\langle C^2 \rangle-\langle C \rangle^2} \, .
\end{equation}
Our task is to find the filter function $\widetilde{Q}(f)$ that maximizes this
signal to noise ratio. A lengthy but straightforward calculation yields
\begin{equation}
{\langle N^2 \rangle} =\frac{T}{4}\int_{-\infty}^{\infty} \vert\widetilde{Q}(f)\vert^2
 M(f)\, df\, .
\end{equation}
where
\begin{eqnarray}
M(f) &=& S_{n1}(f)\, S_{n2}(f) +
S_{n1}(f)S_h(f){\cal R}_2(f)+S_{n2}(f)S_h(f){\cal R}_1(f)\nonumber \\
&& \hspace{1in} +S_h^2(f)\left(\vert\gamma(f)\vert^2+
\frac{{\cal R}_1(f){\cal R}_2(f)}{(2/5\, \sin^2\beta)^2}\right) \, .
\end{eqnarray}
Here $S_{n, i}$ and ${\cal R}_i$ denote the spectral noise and the transfer function
for the $i^{\rm th}$ interferometer. The square of the signal to noise ratio
can be written as
\begin{equation}
{\rm SNR}^2 = \frac{8 T}{25} \sin^4\beta 
\; \frac{\{ \widetilde{P},\widetilde{Q}\}^2 }
{\{ \widetilde{Q}, \widetilde{Q}\} } \, ,
\end{equation}
where $\{ A,B\}$ denotes the inner product\cite{bruce}
\begin{equation}
\{ A,B\} = \int_{-\infty}^\infty df \, A^*(f)  B(f)  M(f) \, ,
\end{equation}
and
\begin{equation}
\widetilde{P}(f)= \frac{S_h(f) \gamma^*(f) }
{M(f) } \, .
\end{equation}
The signal to noise ratio is maximized by choosing the optimal filter ``parallel''
to $\widetilde{P}$. Since the normalization of $\widetilde{Q}$ drops out,
we set $\widetilde{Q}(f)=\widetilde{P}(f)$\cite{flan,bruce}.
Using this filter, the optimal signal to noise ratio for the cross
correlated interferometers is given by
\begin{equation}
    {\rm SNR}^2 = \frac{8 T}{25} \sin^4\beta \int_0^\infty df 
    \frac{\vert\gamma(f)\vert^2 S_h^2(f)}{M(f)} \, ,
\end{equation}
or, equivalently, as
\begin{equation}\label{fullsnr}
    {\rm SNR}^2 = \frac{9 H_0^4 \sin^4\beta}{50 \pi^4}\, T
    \int_0^{\infty} df \frac{\vert \gamma(f)\vert^2\, \Omega_{{\rm gw}}^2(f)}{f^6
    M(f)} \, .
\end{equation}
In the limit that the noise power in each interferometer is very much larger
than the signal power we find
\begin{equation}\label{stand}
    {\rm SNR}^2 = \frac{9 H_0^4 \sin^4\beta}{50 \pi^4}\, T
    \int_0^{\infty} df \frac{\vert \gamma(f)\vert^2\, \Omega_{{\rm gw}}^2(f)}{f^6
    S_{n1}(f) S_{n2}(f)} \, ,
\end{equation}
which recovers the expressions quoted by Flanagan\cite{flan} and Allen\cite{bruce}
when we set $\beta=\pi/2$. The factor of $\sin^4\beta$
was missed in these papers, as the implicit assumption that $\beta=\pi/2$
crept into supposedly general expressions.
The main new ingredient in our expression are the
transfer functions ${\cal T}_i(f)$ that reside in the overlap reduction
function $\gamma(f)$. The transfer functions prevent us from
performing the integral over the two-sphere in equation (\ref{overlap}) in closed form
except at low frequency. However, it
is a simple matter to numerically evaluate $\gamma(f)$ for a given
detector configuration.

If we are trying to detect a weak stochastic background with
a noisy detector, then equation (\ref{stand}) is the appropriate expression to
use for the SNR. However, there may be some range of frequencies
where the signal power dominates the noise power in each detector.
The contribution to the SNR from a clean frequency window of this sort is given by
\begin{equation}
\fl {\rm SNR}^2(f,\Delta f) = T\, \frac{8 \sin^4\beta}{25}
\int_{f-\Delta f/2}^{f+\Delta f /2} df' \, \frac{\vert \gamma(f')\vert^2}
{\left(\vert \gamma(f')\vert^2 +
{\cal R}_1(f'){\cal R}_2(f')/(2/5\, \sin^2\beta)^2\right)} \, .
\end{equation}
The integrand is approximately equal to $1/2$ for all frequencies, so that
\begin{equation}\label{sigdom}
{\rm SNR}(f,\Delta f) \simeq \frac{2}{5}\sin^2\beta \, \sqrt{T \Delta f}  \, .
\end{equation}
This expression provides a useful lower bound for the SNR that can be achieved
for detectors with a clean frequency window of width $\Delta f$
centered at some frequency $f$. For example, the spectral density due to
white dwarf binaries in our galaxy is expected to exceed LISA's noise spectral density
for frequencies in the range $10^{-4} \rightarrow 3\times 10^{-3}$ Hz. To detect
this background at 90\% confidence requires a signal to noise ratio of SNR=1.65,
which can be achieved with less than three hours of integration time $T$.

\subsection{Cross correlating two LISA interferometers}

One possible modification to the current LISA proposal\cite{lppa} would be to fly
six spacecraft instead of three, and thereby form two independent interferometers.
The cost of doing this is considerably less than twice the cost of the current proposal,
and it has the advantage of providing additional redundancy to the mission. How would
we best use a pair of LISA interferometers? If our main concern is getting better
positional information on bright astrophysical sources, then we would fly the two
interferometers far apart, {\it eg.} with one leading and the other trailing the
Earth. However, if we want to maximize the cross correlation then we need the
interferometers to be coincident and coaligned. However, a configuration of this type
is likely to share correlated noise sources, which would defeat the purpose of cross
correlating the interferometers. A better choice is to use a configuration that is coaligned
but not coincident. This can be done by placing six spacecraft at the corners of a
regular hexagon as shown in Figure 5. Notice that interferometers have parallel
arms and that the corner spacecraft are separated by the diameter of the circle.

\begin{figure}[ht]
\vspace*{3.1in}
\includegraphics{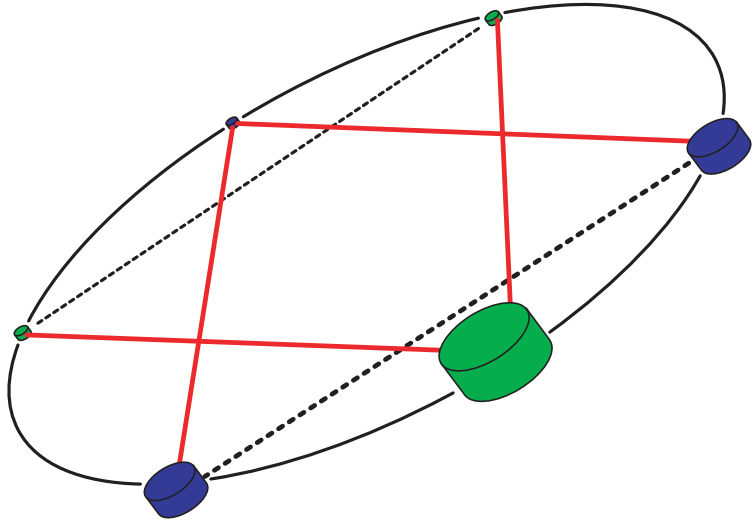}
\caption{The hexagonal cross correlation pattern.}
\end{figure} 

Using the same coordinate system that we used earlier for a single interferometer,
the unit vectors ${\bf u}_i$ and ${\bf v}_i$ along each interferometer arm are:
\begin{eqnarray}
    &&{\bf u}_1 = -{\bf u}_2
    =\frac{1}{2}\hat{x}+\frac{\sqrt{3}}{2}\hat{y} \nonumber \\
    &&{\bf v}_1 = -{\bf v}_2 = -\frac{1}{2}\hat{x}+\frac{\sqrt{3}}{2}\hat{y} \, ,
\end{eqnarray}
where the index $i=1,2$ labels the two interferometers.
The displacement of the corner spacecraft is given by
\begin{equation}
    {\bf x}_1-{\bf x}_2 = - 2 R \, \hat{y}
\end{equation}
where $R$ is the radius of the orbit.
The angle between the interferometer arms is $\beta=\pi/3$, and the length of
each arm is $L=\sqrt{3} R$. The light crossing time between the two
interferometers, $2R/c$, is almost equal to the light crossing time along
each interferometer arm, $\sqrt{3}R/c$. Thus, the loss of sensitivity due to
multiple wavelengths fitting between the interferometers occurs for frequencies
near the transfer frequency $f_*$ (the transfer frequency corresponds to wavelengths
that fit inside the interferometer arms).
For LISA the transfer frequency is $f_*=9.54 \times 10^{-3}$ Hz.

\begin{picture}(0,0)
\put(30,-130){$\gamma(f)$}
\put(210,-265){$\log(f/f_*)$}
\end{picture}
\begin{figure}[ht]
\vspace*{3.6in}
\includegraphics{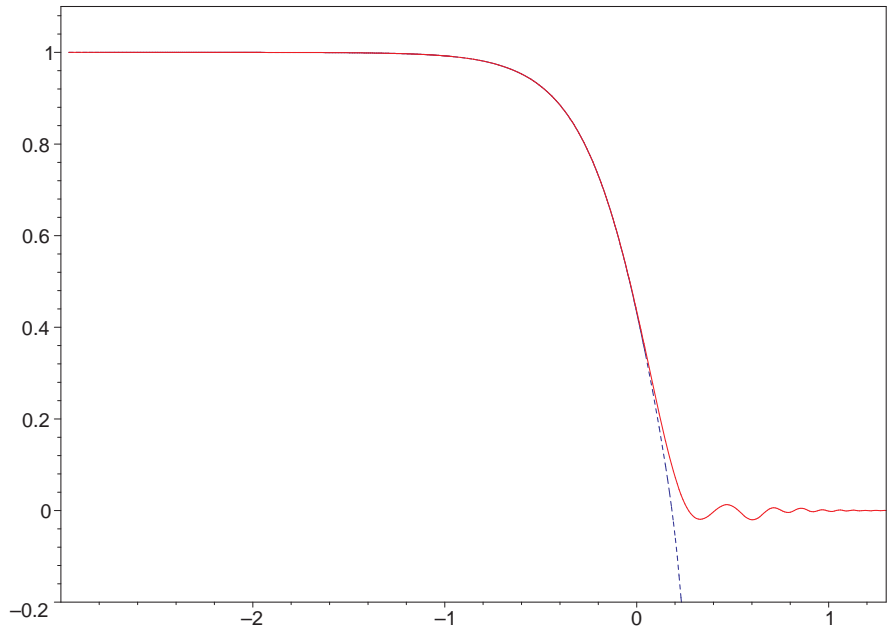}
\caption{The overlap reduction function for the hexagonal cross correlation patern. The
solid line was generated numerically while the dashed line is our analytic
approximation from equation (\ref{gaman}).}
\end{figure} 

The various ingredients
we need to calculate $\gamma(f)$ follow from those given in equations
(\ref{ing1}) through (\ref{ing2}). Putting everything together in (\ref{overlap})
and working in the low frequency limit we find
\begin{equation}\label{gaman}
    \gamma(f) = 1 - \frac{383}{504}\left(\frac{f}{f_*}\right)^2+ 
    \frac{893}{3888}\left(\frac{f}{f_*}\right)^4 -
    \frac{5414989}{143700480}\left(\frac{f}{f_*}\right)^6 + \dots
\end{equation}
For frequencies above $f_*$ the overlap reduction function decays as $f^{-2}$.
A numerically generated plot of $\gamma(f)$ is displayed in Figure 6. Scaling
the signal to noise ratio (\ref{stand}) in units appropriate to a pair of LISA interferometers
we have
\begin{equation}
\fl {\rm SNR}^2 = 34.4 \left(\frac{T}{{\rm year}}\right)\int_0^\infty
\left(\frac{df}{{\rm mHz}}\right)\, \vert \gamma(f)\vert^2 \left(\frac{\Omega_{{\rm gw}}(f)
h_0^2}{10^{-15}}\right)^2 \left(\frac{{\rm mHz}}{f}\right)^6 
\left(\frac{10^{-41}\,{\rm Hz}^{-1}}{S_n(f)}\right)^2 \, .
\end{equation}
Using equation (\ref{lisan}) for $S_n(f)$ and assuming a scale invariant stochastic background
yields a signal to noise of
\begin{equation}
{\rm SNR} = 1.44 \sqrt{\frac{T}{{\rm year}}} \left(\frac{\Omega_{{\rm gw}}
h_0^2}{10^{-14}}\right) \, .
\end{equation}
This represents a 500 fold improvement on the sensitivity of a single LISA detector.
Indeed, if it were not for the astrophysical foregrounds, a pair of LISA detectors
would be well poised to detect a scale invariant gravitational wave
background from inflation.

\section{Missions to detect the CGB}

We will work on the assumption that
astrophysical foregrounds swamp the CGB for frequencies above a few $\mu$Hz,
and design our missions accordingly. What we have in mind is a post-LISA mission
based on the (by then) tried and tested LISA technology, but with some allowance
for improvements in basic technologies such as the accelerometers. Since
cross-correlating two interferometers  at ultra-low frequencies does not buy us a
major improvement in sensitivity, we need to start with a design that has excellent
sensitivity at low frequencies. Basically this means
building bigger interferometers with better accelerometers. To be concrete, we
show the sensitivity curves for three generations of LISA missions in Figure 7.
LISA I corresponds to the current LISA design with the
spacecraft cart-wheeling about the Sun at 1 AU, separated by $L=5\times 10^9$ m.
LISA II refers to a possible follow on mission with the three spacecraft evenly
spaced around an orbit at 1 AU, so that the spacecraft are separated by $L=\sqrt{3}$ AU.
The LISA II mission would use similar optics\footnote{The main difference is that
we have to ``lead our target'' by a much larger amount for the LISA II mission. In other
words, the angle, $\Delta \theta$, between the received and transmitted laser
beams is much larger for LISA II than for LISA I. A simple calculation
yields $\Delta \theta = 2 v/c$, where
$v$ is the velocity around the circle shown in Figure 5. For LISA II
$v/c=(GM_{\odot}/c^2/R)^{1/2}$, while for LISA I $v/c=2 e (GM_{\odot}/c^2/R)^{1/2}$. Here $e=0.01$
is the eccentricity of the LISA I orbits and $R=1$ AU. This equates to lead of
$\Delta \theta = 4\times 10^{-6}$ radians for LISA I and $\Delta \theta = 2\times 10^{-4}$
radians for LISA II.} to LISA I (same laser power and telescope size), but allows for
an order of magnitude improvement in accelerometer performance. LISA III is similar to LISA II,
except that the constellation would orbit at 35 AU (between Neptune and Pluto) and
the accelerometers would be improved by a further two orders of magnitude. The
acceleration noise for LISA III would benefit from the three orders of magnitude reduction
in solar radiation relative to LISAs I and II, but this would come at the cost of
having to power the spacecraft using nuclear generators (RTGs).

\begin{figure}[ht]
\vspace*{3.5in}
\includegraphics{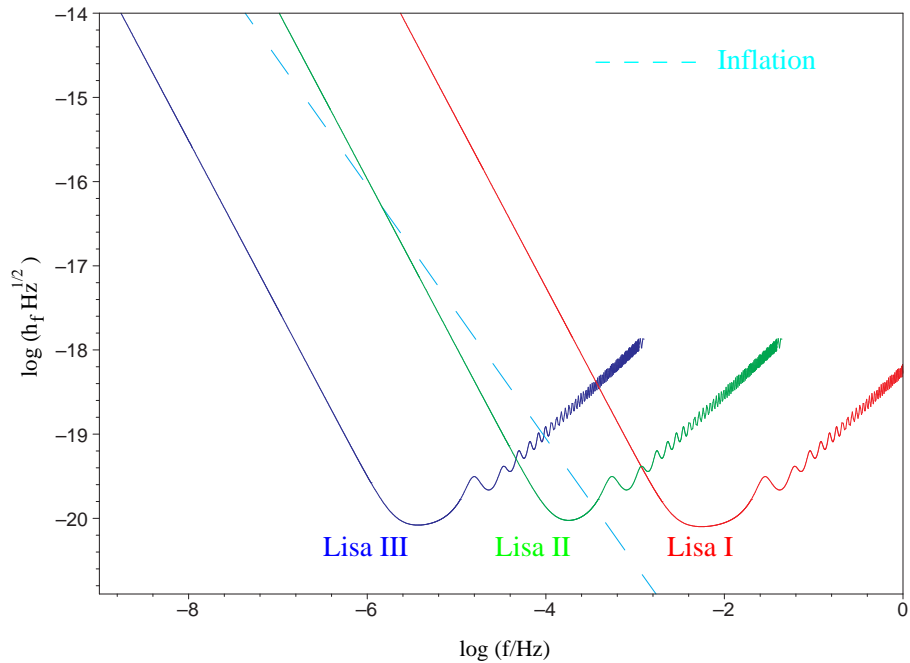}
\caption{Sensitivity curves for three generations of LISA missions. Also shown
is a prediction for the CGB in a scale invariant inflationary
scenario.}
\end{figure} 

For all three missions we see that the $\mu$Hz range lies well below the interferometer's
transfer frequency
\begin{equation}
f_* = 1.84 \times 10^{-4} \left(\frac{\sqrt{3} \, {\rm
    AU}}{L}\right) \; {\rm Hz}\, .
\end{equation}
Indeed, it would take a mission orbiting at 180 AU (
somewhere inside the Kuiper belt) to achieve a transfer frequency of $f_*=10^{-6}$ Hz.
In practical terms this allows us to work in the low frequency limit $f \ll f_*$ where
it is easy to derive analytic expressions for the overlap reduction function $\gamma(f)$.
Moreover, we need only consider acceleration noise when working below 1$\mu$Hz. All our
calculations are based on the same hexagonal cross correlation shown in Figure 5 that
we used for the LISA I cross correlation. 
Using equation (\ref{accel}) for the acceleration noise and ignoring the position noise yields
\begin{equation}\label{an}
S_n(f) = 3.42 \times 10^{-33} \left(\frac{\sqrt{3} \, {\rm
    AU}}{L}\right)^2 \left(\frac{\mu{\rm Hz}}{f}\right)^4 
    \left(\frac{\delta a}{3\times 10^{-16} {\rm m}{\rm
    s}^{-2}}\right)^2 \, {\rm Hz}^{-1}.
\end{equation}
In the limit that the noise power dominates the spectral density of the CGB we arrive at
an estimate for the signal to noise ratio for frequencies below 1$\mu$Hz
\begin{eqnarray}
   {\rm SNR}^2 &=& 29.4 \left(\frac{T}{{\rm
    year}}\right)\left(\frac{L}{\sqrt{3} \, {\rm
    AU}}\right)^4 \left(\frac{3\times 10^{-16} {\rm m}{\rm
    s}^{-2}}{\delta a}\right)^4 \times \nonumber \\ \bs
    && \quad 
    \int_0^{f} \vert \gamma(f')\vert^2 \left(\frac{\Omega_{\rm gw}(f')
    h_{0}^{2}}{10^{-14}}\right)^2
    \left(\frac{f'}{\mu{\rm Hz}}\right)^2 \left(\frac{df'}{\mu{\rm
    Hz}}\right) \, .
\end{eqnarray}
The SNR is scaled against the LISA II specifications. Since $f/f_* \ll 1$
it is a good approximation to set $\gamma(f)=1$ inside the integral. We model
the CGB spectrum near 1$\mu$Hz by a simple power law:
\begin{equation}
\Omega_{\rm gw}(f) = \Omega_{\mu{\rm Hz}}\left(\frac{f}{\mu{\rm Hz}}\right)^\alpha \, ,
\end{equation}
scaled relative to a reference value at 1$\mu$Hz. With these approximations
we arrive at our final expression for the SNR for a pair of cross correlated interferometers
in the ultra low frequency regime:
\begin{eqnarray}\label{snr}
 && \fl {\rm SNR} = \frac{3.13}{\sqrt{1+2\alpha/3}} \left(\frac{T}{{\rm
    year}}\right)^{1/2}\left(\frac{L}{\sqrt{3} \, {\rm
    AU}}\right)^2 \left(\frac{3\times 10^{-16} {\rm m}{\rm
    s}^{-2}}{\delta a}\right)^2\left(\frac{\Omega_{\mu{\rm Hz}}\,
    h_{0}^{2}}{10^{-14}}\right)\left(\frac{f}{\mu{\rm Hz}}\right)^{\alpha+3/2} \nonumber \\
    && \,
\end{eqnarray}
We can compare this to the SNR of a single interferometer in the ultra low frequency regime
by combining equations (\ref{ogw}), (\ref{snrf}) and (\ref{an}) to find
\begin{equation}
 \fl  {\rm SNR}_s(f) = 0.683 \left(\frac{L}{\sqrt{3} \, {\rm
    AU}}\right)^2 \left(\frac{3\times 10^{-16} {\rm m}{\rm
    s}^{-2}}{\delta a}\right)^2\left(\frac{\Omega_{\mu{\rm Hz}}\,
    h_{0}^{2}}{10^{-14}}\right)\left(\frac{f}{\mu{\rm Hz}}\right)^{\alpha+1} \, .
\end{equation}
We see that the individual and cross correlated sensitivities are related:
\begin{equation}
{\rm SNR} = \frac{0.82}{\sqrt{1+2\alpha/3}}\, (T \Delta f)^{1/2}\;
{\rm SNR}_s(f) \, , \quad {\rm for} \; f \approx \Delta f \approx 1\, \mu{\rm Hz}\, .
\end{equation}
This supports our earlier assertion that cross correlating two interferometers improves
on the sensitivity of a single interferometer by $(T\Delta f)^{1/2}$ where $\Delta f$ is
approximately equal to the central observing frequency $f$.

Equation (\ref{snr}) tells us that a GABI detector built from two LISA II interferometers
could detect the CGB at greater than 90\% confidence
with one year of data taking if $\Omega_{\rm gw}(1\mu{\rm Hz})h_{0}^2 = 10^{-14}$. The same
equation also tells us what has to be done to achieve greater sensitivity. It is
clear that increasing the duration of the mission is not the best answer as the SNR only
improves as the square root of the observation time. 
In contrast, increasing the size of
the interferometer or reducing the acceleration noise produces a quadratic increase in
sensitivity. For example, a GABI detector built from
a pair of LISA III interferometers could detect the CGB at 90\% confidence
for signals as small as $\Omega_{\rm gw}(1\mu{\rm Hz})h_{0}^2 = 4\times 10^{-22}$. In-fact,
for signal strengths above this value the LISA III detectors are signal dominated
and we need to use equation (\ref{sigdom}) instead of (\ref{snr}) to calculate the
SNR.

We can also use equation (\ref{snr}) to determine how well a GABI mission can measure
the CGB power spectrum. Suppose that we break up the frequency spectrum into bins of width
$\delta f$. Nyquist's theorem tells us that $\delta f \geq 1/T$, {\it eg.} for an
observation time of one year the frequency resolution is $3.17\times 10^{-8}$ Hz.
Writing $\delta f = n/T$ where $n\geq 1$, and taking $\delta f \ll f$, we find from (\ref{snr}) that
in the frequency window $(f-\delta f/2,f+\delta f /2)$ the SNR is
\begin{equation}
\fl {\rm SNR}(f,\delta f) = 0.965 \sqrt{n} \left(\frac{L}{\sqrt{3} \, {\rm
    AU}}\right)^2 \left(\frac{3\times 10^{-16} {\rm m}{\rm
    s}^{-2}}{\delta a}\right)^2\left(\frac{\Omega_{\mu{\rm Hz}}\,
    h_{0}^{2}}{10^{-14}}\right)\left(\frac{f}{\mu{\rm Hz}}\right)^{\alpha+1} \, .
\end{equation}
We see that the SNR in each frequency bin scales as $f^{1+\alpha}$.
For a scale invariant spectrum ($\alpha = 0$) this translates into poor
performance at frequencies below 1 $\mu$Hz and limits the range over
which we can measure the spectrum. For example, if the CGB has a scale invariant spectrum and an
amplitude of $\Omega_{{\rm gw}}h_0^2 \geq 2 \times 10^{-13}$,
a pair of LISA II interferometers could measure the spectrum over the range
$10^{-7} \rightarrow 10^{-6}$ Hz. For a pair of LISA III detectors the main limitation
at low frequencies comes from Nyquist's theorem. Since mission lifetimes are limited to tens of years,
it will not be possible to measure the CGB spectrum much below $10^{-8}$ Hz using space based
interferometers. Indeed, it may be difficult to push much below $10^{-7}$ Hz unless ways can be
found to build detectors that are stable for many months. A pair of LISA III interferometers
could measure the spectrum between
$10^{-8} \rightarrow 10^{-6}$ Hz if $\Omega_{{\rm gw}}(10^{-8}\, {\rm Hz})
h_0^2 \geq 10^{-18}$. To measure the CGB spectrum below $10^{-8}$ Hz requires a return, full circle,
to the world of CMB physics. Detailed polarization measurements of the CMB can be used to infer\cite{rob}
the CGB power spectrum for frequencies in the range $10^{-18} \rightarrow 10^{-16}$ Hz.

\section{Discussion}

The LISA follow-on missions we have described will be able to detect or place
stringent bounds on the CGB amplitude and spectrum between $10^{-8}$ and $10^{-6}$ Hz.
But how can we be sure that it is the CGB we have detected and not some unresolved
astrophysical foreground? The answer can be found in the statistical character of the competing
signals. Most early universe theories predict that the  CGB is truly stochastic. In contrast,
the astrophysical signal is only approximately stochastic, in a sense that can be made precise by
appealing to the central limit theorem. The line in Figure 1 indicating the predicted amplitude
of various astrophysical foregrounds corresponds to what is known as
the confusion limit. It marks the amplitude at which we can expect to find, on average, one source
per frequency bin\footnote{The confusion limit goes down when the frequency resolution goes up.
Most plots of the confusion limit assume a one year observation period so the bins are
$3.17\times 10^{-8}$ Hz in width.}. It is not until we reach signal strengths considerably smaller
than the confusion limit that the astrophysical signal starts to look stochastic. And therein lies
the answer to our question. A ``stochastic'' signal due to astrophysical sources will always have
bright outliers that sit just above the amplitude of the stochastic signal. The number
and distribution of the outliers can be predicted on statistical grounds.
If we do not see the outliers, then we can safely conclude that we have detected the CGB.

\section*{Acknowledgements}

We thank Peter Bender and Raffaella Schneider for sharing their thoughts on possible
astrophysical sources of gravitational waves.

\end{document}